\documentclass[twocolumn,showpacs]{revtex4}

\def\hc2{$H_{c2}$}

\def\cuscn{$\kappa$-(BEDT-TTF)$_2$Cu(NCS)$_2$}

\def\nh4{$\alpha$-(BEDT-TTF)$_2$NH$_4$Hg(SCN)$_4$}

\newcommand{\ltsim}{\mbox{{\raisebox{-0.4ex}{$\stackrel{<}{{\scriptstyle\sim}}
$}}}}

\usepackage{graphicx}
\usepackage{dcolumn}
\usepackage{amsmath}
\usepackage{longtable}
\begin{document}

\title{Comparison of the Fermi-surface topologies of
$\kappa$-(BEDT-TTF)$_2$Cu(NCS)$_2$ and its deuterated analogue}

\author{R.S.~Edwards$^1$, A. Narduzzo$^1$, J. Singleton$^{1,2}$, A. Ardavan$^1$
and J.A. Schlueter$^3$}

\affiliation{$^1$Oxford University Department
of Physics, Clarendon Laboratory, Parks Road, Oxford OX1 3PU, United Kingdom.\\
$^2$National High Magnetic Field Laboratory,
Los Alamos National Laboratory, TA-35, MS-E536,
Los Alamos, NM87545, USA\\
$^3$Materials Science Division,
Argonne National Laboratory, Argonne, Illinois 60439, USA}

\begin{abstract}
We have measured details of the quasi one-dimensional
Fermi-surface sections in the organic
superconductor $\kappa$-(BEDT-TTF)$_2$Cu(NCS)$_2$ and
its deuterated analogue using angle-dependent millimetre-wave
techniques. There are significant differences in the
corrugations of the Fermi surfaces in the deuterated and
undeuterated salts. We suggest that this is important
in understanding the inverse isotope effect, where the
superconducting transition temperature rises on deuteration.
The data support models for
superconductivity which invoke electron-electron
interactions depending
on the topological properties
of the Fermi surface.
\end{abstract}

\pacs{71.18.+y, 71.27.+a, 72.80.Le, 74.70.-b, 78.70.Gq}

\maketitle
\date{today}

The nature of the superconducting groundstate
in quasi-two-dimensional charge-transfer salts
such as \cuscn ~has attracted much recent experimental~\cite{sasaki,elsinger,hillspat,biggs}
and theoretical~\cite{schmalian,kuroki,charffi,aoki,lee} interest.
The majority of the experimental data suggest that the superconductivity
is not describable by a simple BCS-like, phonon-mediated approach
(for a review, see Refs.~\cite{biggs,jschm} and refs. therein).
Consequently, a number of the theoretical
treatments invoke pairing mediated by
electron-electron interactions and/or antiferromagnetic
fluctuations~\cite{schmalian,kuroki,charffi,aoki,lee}.
In such a scenario, the ``nestability'' of the Fermi surface
is an important consideration; it
is expected that alterations of the Fermi-surface
topology will affect the superconducting
transition temperature~\cite{schmalian,kuroki,charffi,aoki,lee}.

In this context, the observation of a
``negative isotope effect'' in
\cuscn ~may be of great importance~\cite{kini,schlueter};
on replacing the
terminal hydrogens of the BEDT-TTF molecule
in \cuscn ~by deuterium, it was found that
an increase ($\Delta T_{\rm c} \approx 0.3$~K) in the
superconducting critical temperature
$T_{\rm c}$ occurred~\cite{kini,schlueter}.
By contrast, isotopic substitutions of other
atoms in the BEDT-TTF molecule or in the anion layer produce a
very small, normal isotope effect or no significant isotope
effect at all, respectively~\cite{schlueter}.
In this paper we describe millimetre-wave measurements
which compare the Fermi surfaces of deuterated
and conventional samples of \cuscn .
~The data suggest that
it is primarily the changes in the topology
of the Fermi surface
brought about by deuteration that cause the observed
isotope effect, supporting models for superconductivity
involving pairing via electron-electron
interactions~\cite{schmalian,kuroki,charffi,aoki,lee}.

The Fermi surface of \cuscn
~comprises a quasi-two-dimensional (Q2D) pocket
(the $\alpha$ pocket) and a pair of
quasi-one-dimensional (Q1D) sheets~\cite{goddard}; it
is similar to that used by Pippard to predict magnetic
breakdown~\cite{pippard}.
It is known that the $\alpha$ pocket is
hardly affected by deuteration~\cite{biggs}, and so
our measurement concentrates on the Q1D sheets.
Fermi-surface traversal resonances (FTRs)~\cite{me+mar5,ardavan}
(i.e. resonances in the high-frequency conductivity
caused by magnetic-field-induced motion of quasiparticles
across the Fermi sheets) are used to infer the
corrugations of the sheets.

\begin{figure}[htbp]
\centering
\includegraphics[height=10cm]{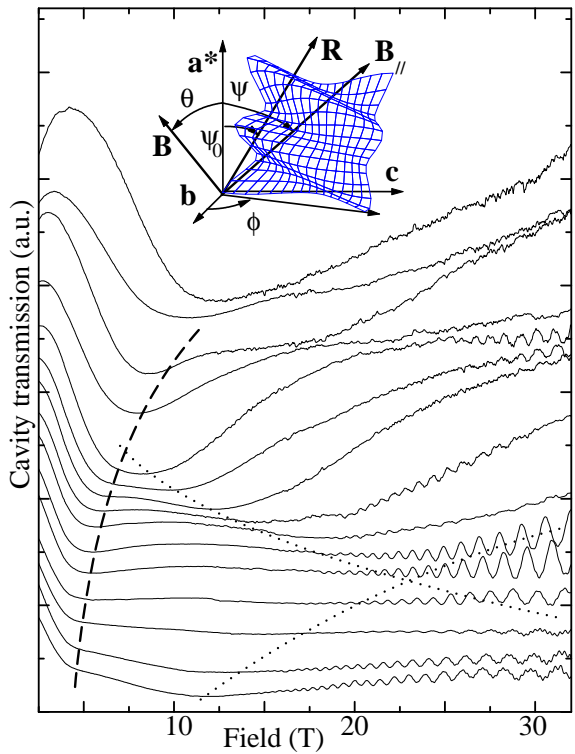}
\centering \caption{Transmission of a cavity loaded with a
d8 \cuscn ~sample versus magnetic field. The temperature was 1.5~K and
$\phi = 5^\circ$. Field sweeps at angles from
$\theta = 0^\circ$ (bottom) to $70^\circ$ (top) in $5^\circ$ steps
are shown, normalised and offset for clarity.
The thick dashed line shows the position of a feature associated with the
superconducting-to-normal transition, while the fine dashed lines show
the FTRs.
Inset: the relationship between the coordinates
{\bf B}, $\theta$ and $\phi$, the component of the field ${\bf B}_{||}$
and the angle $\psi$. {\bf R} indicates the
axis of the Fermi-surface corrugation.
} \label{data}
\end{figure}

The experiments involved single crystals of
\cuscn  ~($\sim 0.7 \times 0.5 \times 0.1$~mm$^3$; mosaic spread
$\ltsim 0.1^{\circ}$), produced using
electrocrystallization~\cite{kini,schlueter}.
In some of the crystals, the terminal hydrogens of the BEDT-TTF molecules
were isotopically substituted by deuterium;
we refer to the deuterated samples as d8, and
conventional hydrogenated samples as h8.
A single sample is mounted at the centre
(in a magnetic field antinode)
of a rectangular cavity
of inner dimensions $1.55 \times 3.10 \times 6.00$\,mm$^3$
resonating at 72~GHz in the $TE_{102}$ mode~\cite{me+mar5};
the oscillating $H$-field lies within the sample's Q2D ({\bf b},~{\bf c}) planes.
In this configuration, the effective skin depth is very large,
and the GHz fields penetrate the bulk of the sample~\cite{me+mar5}.
The cavity can be rotated with respect to the external
quasistatic magnetic field {\bf B} so as to vary the angle $\theta$ between
{\bf B} and the normal to the sample's Q2D planes~\cite{me+mar5};
the normal to the Q2D planes is the ${\bf a}^*$ direction of
the reciprocal lattice~\cite{crystal,watanabe}.
In addition, the sample can be turned about ${\bf a}^*$
within the cavity,
so as to vary the plane of rotation, defined by the
azimuthal angle $\phi$~\cite{me+mar5}.
The angles $\theta$ and $\phi$ and their relationship to the
Q1D sheet of the Fermi surface are given in the inset to
Fig.~\ref{data}.

Experiments were carried out on samples of  d8 \cuscn
for angles $-70^{\circ} \leq \theta \leq 70^{\circ}$
for four different azimuthal angles $\phi$.
Fig.~\ref{data} shows results for an azimuthal angle of
$\phi=5^\circ$ for $\theta$ values between $0^\circ$ and $70^\circ$ in
$5^\circ$ steps, at a temperature of 1.5~K.
At low fields (around 4~T at $\theta = 0^\circ$) one absorption can be
seen (thick dashed line in Fig.~\ref{data}). This is related to the
superconducting to normal transition of the sample~\cite{hillspat};
it follows the $\theta$ dependence of the upper
critical field, $\mu_0 H_{\rm c2}$, which varies approximately
as $1/\cos\theta$~\cite{msnam5}. At high field magnetic quantum oscillations
are observed, indicating that the sample is pure;
the angular behaviour of the frequency $F$ of
the oscillations ($F \propto 1/\cos\theta$)
provides a check of the angle $\theta$~\cite{jschm}.
At some $\theta$, magnetic breakdown oscillations,
caused by tunnelling between the
Q2D and Q1D Fermi-surface sections, are superimposed on the
lower frequency oscillations caused by the Q2D pocket~\cite{jschm,me+mar5}.
At intermediate fields there are two broad absorptions~\cite{steve1}
(fine dashed lines); their $(\theta,\phi)$
dependence (see below) allows them to be
unambigously attributed to FTRs caused by the Q1D sheets.

The field positions of the FTRs were
recorded for all angles studied.
In order to analyse the FTRs, the experimental coordinates
{\bf B}, $\theta$ and $\phi$ must be converted into
the component of the field ${\bf B}_{||}$
within the plane of the Q1D Fermi-surface sheets,
and the angle $\psi$ between the normal to the sample's
Q2D planes and ${\bf B}_{||}$ (see Fig.~\ref{data}, inset)~\cite{me+mar5}.
This done via the equations~\cite{me+mar5}
\begin{eqnarray}
B_{||} &=& B \sqrt{\sin^2\theta\cos^2\phi +
\cos^2\theta},\nonumber \\ \tan\psi &=& \tan\theta\cos\phi.
\label{ftreqns}
\end{eqnarray}
Each corrugation of the Q1D Fermi sheets is expected to give
rise to a FTR with the $\psi$ dependence~\cite{me+mar5,ftrtheory},
\begin{equation}
\frac{\omega}{B_{||}}=A\sin(\psi-\psi_0).
\label{boink}
\end{equation}
Here, $\omega$ is the angular frequency of the
millimetre-waves, $A$ is a constant depending on details of the
Fermi surface~\cite{me+mar5}, and $\psi_0$ defines the axis of
the corrugation ${\bf R}$ (Fig.~\ref{data}, inset).
As the millimetre-wave frequency is held constant,
the FTRs should lie on sinusoidal ``arches''
when $1/B_{||}$ is plotted as a function of $\psi$~\cite{me+mar5}.

\begin{figure}[htbp]
\centering
\includegraphics[height=7cm]{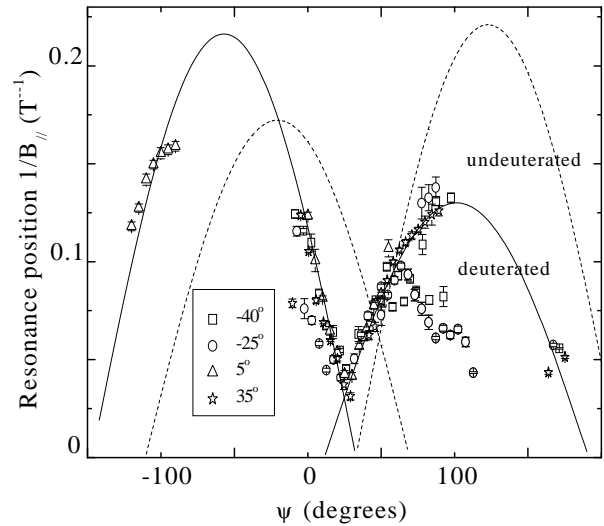}
\centering \caption{The field positions of the resonances in
d8 \cuscn ~in FTR coordinates. Data for four different
azimuthal angles studied at 1.5~K are shown; squares $\phi=-40^\circ$;
circles $\phi=-25^\circ$; triangles $\phi=5^\circ$; stars
$\phi=35^\circ$. Two curves (solid lines) show the fits to the
resonances using Eqn~\ref{boink}.
The dotted lines show equivalent fits to data from h8 \cuscn.
For clarity the data points have been omitted (see Ref.~\cite{me+mar5}
for representative data).} \label{d8ftr}
\end{figure}

Fig.~\ref{d8ftr} shows the FTR positions plotted in
terms of $1/B_{||}$ and $\psi$.
Apart from a region close to $\psi = +100^{\circ}$
where the feature associated with the superconducting to normal
transition obscures the FTRs at some $\phi$,
making the exact position difficult to gauge,
the data lie on two ``arches'', shown as curves (Fig.~\ref{d8ftr});
the curves were obtained by fitting the data to Eqn.~\ref{boink}.
This indicates that the Q1D Fermi surface of d8 \cuscn
~has two distinct corrugations,
with their axes ${\bf R}_1$ and ${\bf R}_2$
at angles $\psi_0=17.9^\circ \pm 2.0^\circ$ and
$39.8^\circ \pm 2.0^\circ$ to ${\bf a}^*$ respectively.

Equivalent experiments were carried out in h8 \cuscn .
~(Some representative data are shown in Ref.~\cite{me+mar5}).
Again, the resonance positions lie on two ``arches'' (shown in Fig.~\ref{d8ftr}),
implying that the Q1D
Fermi surface of h8 \cuscn ~has two distinct corrugations~\cite{me+mar5}.
In this case, the corrugation axes ${\bf R}_1$ and ${\bf R}_2$
are at angles $\psi_0 =21.2^\circ \pm 2.0^\circ$
and $-20.8^\circ \pm 2.0^\circ$ to ${\bf a}^*$ respectively.
The data for both samples are summarised in Table~\ref{table1}.

\cuscn ~has a monoclinic crystal
structure with the crystallographic {\bf a}-axis at an angle of
20.3$^\circ$ to the normal to the Q2D planes (${\bf a}^*$)~\cite{crystal};
in this respect, the crystal structures of d8 and h8 \cuscn ~appear
identical~\cite{watanabe}.
Within tight-binding bandstructure, the corrugation axes of
a Fermi surface usually relate to the primitive lattice
translation vectors of the {\it real-space} lattice~\cite{ashcroft}.
Rather than work in terms of the
angle $\psi_0$, which defines the directions of the corrugation axes with respect
to ${\bf a}^*$, it is more useful to use the angle $\Psi_0$,
which relates to the real-space
vector {\bf a}.
Once this is done (Table~\ref{table1}),
it is plain that the corrugation axis ${\bf R}_1$
in both the d8 and h8 samples lies very close to the {\bf a}
(interlayer) direction.
By contrast, the direction of ${\bf R}_2$,
the second corrugation axis, differs;
with reference to the primitive lattice translation vectors
\begin{equation}
{\bf T}_{mn}= m{\bf a} + n{\bf c}
\label{vec}
\end{equation}
where $m$ and $n$ are integers, we find that
in d8, ${\bf R}_2$ is very close in direction to ${\bf T}_{2-1}$,
whereas in h8, it is close in direction to ${\bf T}_{11}$ (Table~\ref{table1}).
The reasons for the dominance of these particular directions are unclear;
however, interlayer coupling through the anion layer
is presently poorly
understood at a molecular-orbital level. It is possible
that a variety of overlap-pathways may be operative and
that the choice of dominant overlap-pathway through the
anion layer depends very sensitively on the exact coordinates
of the terminal
end of the BEDT-TTF molecule.
In this context, it will be very useful to have high-resolution
structural experiments which address the detailed
differences between h8 and d8 at low temperatures~\cite{watanabe}.

\begin{table*}
\begin{ruledtabular}
\begin{tabular}{ccccc}
 & d8 ${\bf R}_1$ & d8 ${\bf R}_2$ & h8 ${\bf R}_1$ & h8 ${\bf R}_2$ \\ \hline
$A/\omega$ & $0.198 \pm 0.004$ & $0.131 \pm .002$ & $0.204 \pm 0.004$ & $0.168 \pm 0.004$ \\
$\psi_0$ & $17.9 \pm 2.0^{\circ}$ & $39.8 \pm 2.0^{\circ}$ & $21.2 \pm 2.0^{\circ}$
& $-20.8 \pm 2.0^{\circ}$ \\
$\Psi_0$ & $-2.4 \pm 2.0^{\circ}$ & $19.5 \pm 2.0^{\circ}$ & $0.9 \pm 2.0^{\circ}$
& $-41.1 \pm 2.0^{\circ}$ \\
${\bf T}_{mn}$ & ${\bf T}_{10}$ & ${\bf T}_{2 -1}$ & ${\bf T}_{10}$ & ${\bf T}_{11}$ \\
$I$ & 1 & $0.34 \pm 0.06$ & 1 & $6 \pm 2$ \\
\end{tabular}
\end{ruledtabular}
\caption{The values for $A/\omega$ (see Eqn.~\ref{boink}), $\psi_0$
(angle of corrugation axis with respect to ${\bf a}^*$)
and $\Psi_0$ (angle of corrugation axis with respect to ${\bf a}$)
for each of the FTRs seen in d8 and h8 \cuscn.
~Also shown are the vectors ${\bf T}_{mn}$ which define the
directions of the corrugation axes ${\bf R_1}$ and ${\bf R}_2$
(see Eqn.~\ref{vec}). $I$ is the average intensity of the FTR
at $\theta = 0^\circ$ normalised as
described in the text.} \label{table1}
\end{table*}

Finally, it is interesting to work out the relative amplitudes of the
corrugations in h8 and d8 \cuscn.
~Models of FTR allow one to relate the intensity of the FTR
to the amplitude of the Fermi-surface corrugation~\cite{ftrtheory}.
Measurements of the d.c. transport properties of d8 and h8
\cuscn ~suggest that the transfer integral $t_{\perp}$
in the {\bf a} direction is very similar in the two materials
($t_\perp \approx 0.04$~meV)~\cite{goddard}.
This implies that the Fermi-surface
corrugations along ${\bf R}_1$ should be very similar in
d8 and h8~\cite{goddard}; in d8 and h8 samples with equal volume the corresponding
FTRs should have the same intensity~\cite{ftrtheory}.
The average intensities $I$ of each FTR for $\theta=0^\circ$ are shown
in Table \ref{table1}. As the samples are of different size, the
intensities of the FTRs have been normalised
to that of the FTR corresponding to ${\bf R}_1$.

\begin{figure}[htbp]
\centering
\includegraphics[height=9cm]{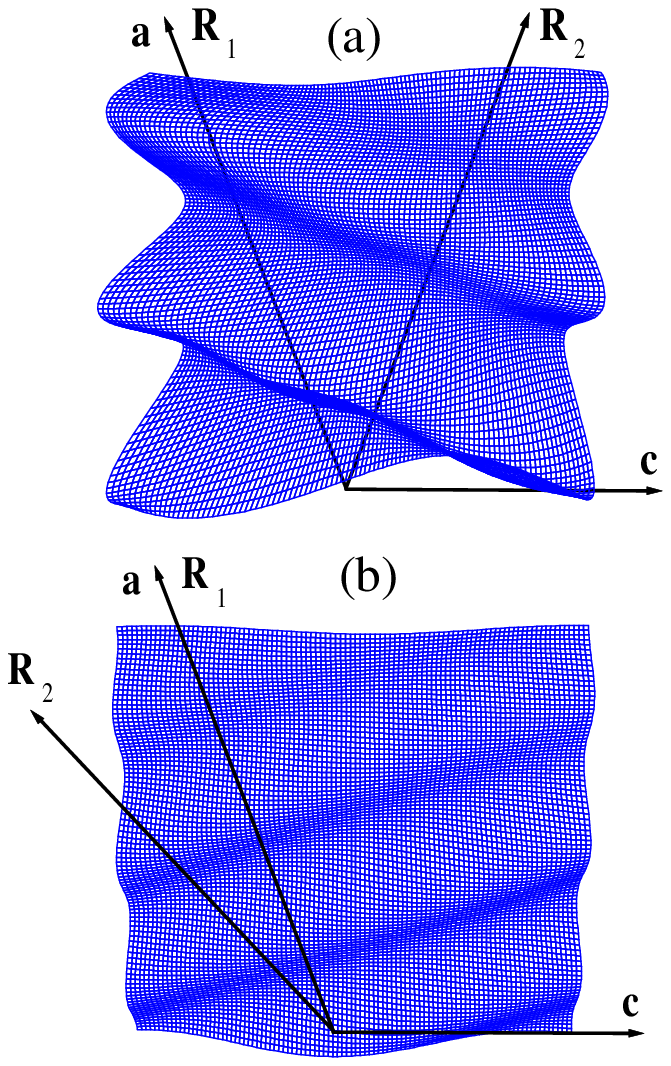}
\includegraphics[height=3cm]{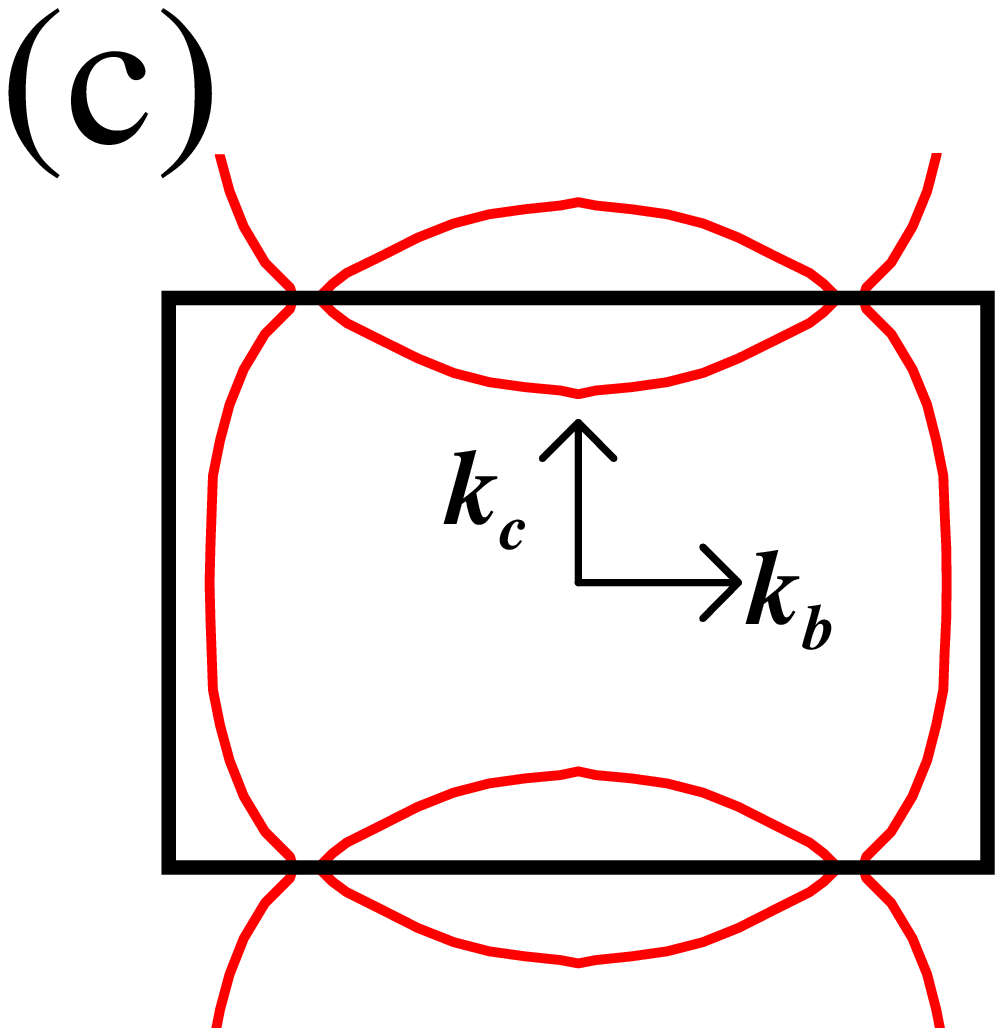}
\centering \caption{(a) Representation of
the Q1D Fermi-sheet
topology in h8 \cuscn
~derived from the fits to the FTR data; (b) the
same for d8 \cuscn ~plotted at the same scale.
The corrugations have been greatly enhanced
for clarity.
(c) Plan view of the Fermi surface of \cuscn, ~showing
the Q2D pocket and Q1D sheets~\cite{goddard};
(a) and (b) represent side views ({\it i.e.} looking along ${\bf k_b}$)
of the Q1D sheets.} \label{fs}
\end{figure}

Using the periodicity in $k$-space~\cite{crystal}, the relative intensities
of the FTRs, and
the orientations $\Psi_0$ of ${\bf R}_1$ and ${\bf R_2}$,
it is possible to make
a comparison of the Q1D Fermi sheets for both materials. Fig.
\ref{fs} shows these representations, with the corrugations,
assumed sinusoidal,
shown at the same scale. This scale is chosen so that the
differences between d8 and h8 are clear; in
reality, the small measured value of $t_{\perp}$~\cite{goddard}
suggests that the corrugations will be on an extremely small scale.

~

Despite an intensive search,
no features attributable to cyclotron resonance (CR) due to
the Q2D Fermi-surface pocket were observed in either d8 or h8, in agreement
with previous studies~\cite{me+mar5}.
The response of our cavity system is dominated by
the interlayer component of the sample's high-frequency
conductivity~\cite{me+mar5,ardavan}.
Fermi-surface sections with more complex corrugations
in the interlayer direction will dominate the high-frequency
interlayer conductivity~\cite{ftrtheory};
recent resistivity measurements of \cuscn ~suggest that
the corrugations of the Q2D pocket are simpler
and more regular than those of the Q1D sheets~\cite{goddard},
perhaps explaining the absence of CR.
Similarly, the {\it in-plane} corrugations of the Q1D sheet (Fig.~\ref{fs}(c))
will have little effect on the high-frequency
{\it interlayer} conductivity~\cite{ftrtheory},
and therefore do not result in detectable FTRs.

It is obvious that there is a difference between the Q1D
Fermi sheets of the two materials, with the corrugations in
h8 being stronger; the dominant corrugation has axis ${\bf R}_2$, at
-41.1$^\circ$ to {\bf a}. By contrast, the corrugations
in d8 \cuscn ~are weaker, and are dominated by that with axis
${\bf R}_1$ lying along
{\bf a}. This suggests that the Fermi surface of
d8 \cuscn ~would be more amenable to nesting than
that of h8.

Our data support models for superconductivity such as those of
Refs.~\cite{kuroki,charffi}. In these, the
pairing of electrons is mediated by electron-electron interactions
which depend on the ``nestability'' of the
Fermi-surface; hence they predict a $T_{\rm c}$ which is
sensitive to the details of the Fermi-surface topology.
The difference between the Q1D Fermi sheets of d8
and h8 \cuscn ~measured using FTR can thus explain the isotope effect;
the Q1D Fermi sheets in the d8 samples are less
corrugated (and therefore more nestable), leading to a higher
$T_{\rm c}$.

In summary, we have measured details of the Fermi-surface topology
of the deuterated organic superconductor \cuscn , ~and compared
them with equivalent measurements of the undeuterated salt.
We find that the quasi-one-dimensional Fermi-surface
sheets are significantly more corrugated in the undeuterated salt,
perhaps explaining the ``inverse isotope effect'' observed on deuteration.
Our data support models for
exotic d-wave superconductivity in the
organics which invoke electron-electron
interactions depending
on the topological properties
of the Fermi surface.

This work is supported by EPSRC (UK).
NHMFL is supported by the
US Department of Energy (DoE), the National
Science Foundation and the State of Florida.
Work at Argonne is sponsored
by the DoE, Office of Basic Energy Sciences,
Division of Materials Science under contract number
W-31-109-ENG-38.
We thank Stephen Hill and Stephen Blundell for
constructive comments.


\begin{thebibliography}{Bibliography}
\bibitem{steve1}
For most $\theta$, these absorptions occur
well away from the regions
of significant magnetic breakdown.
Hence, the quasiparticle motion across the Q1D
sheets is decoupled from that on the Q2D pocket;
resonances can be unambiguously associated with either
the Q1D or Q2D sections of the Fermi surface.
\end{thebibliography}
\end{document}